# ENERGY EFFICIENT WIRELESS COMMUNICATION USING GENETIC ALGORITHM GUIDED FASTER LIGHT WEIGHT DIGITAL SIGNATURE ALGORITHM (GADSA)

Arindam Sarkar[1] and J. K. Mandal[2]

[1]Department of Computer Science & Engineering, University of Kalyani, W.B, India
arindam.vb@gmail.com
[2]Department of Computer Science & Engineering, University of Kalyani, W.B, India
jkm.cse@gmail.com

## ABSTRACT

*In this paper GA based light weight faster version of Digital Signature Algorithm (GADSA) in wireless communication has been proposed. Various genetic operators like crossover and mutation are used to optimizing amount of modular multiplication. Roulette Wheel selection mechanism helps to select best chromosome which in turn helps in faster computation and minimizes the time requirements for DSA. Minimization of number of modular multiplication itself a NP-hard problem that means there is no polynomial time deterministic algorithm for this purpose. This paper deals with this problem using GA based optimization algorithm for minimization of the modular multiplication. Proposed GADSA initiates with an initial population comprises of set of valid and complete set of individuals. Some operators are used to generate feasible valid offspring from the existing one. Among several exponents the best solution reached by GADSA is compared with some of the existing techniques. Extensive simulations shows competitive results for the proposed GADSA.*

## 1. INTRODUCTION

Digital Signature Algorithm [1, 2] performs modular exponentiation for encryption/decryption purpose. This DSA comprises of a public key, private key and modulus as given in equation (1).

$$c = p^E \bmod N \qquad\qquad (1)$$

Where plain text (p) is a positive integer in the range of [0, 1, 2, 3, N-1] in DSA technique. N is a result of multiplication of two very large prime number and E is a randomly chosen positive integer which satisfies this equation. In this exponentiation large numbers of multiplications are requisite. In contemporary cryptographic technique E value could be greater than 128 bits [6]. For this explanation in asymmetric key cryptography encryption as well as decryption is very computationally costly.

This work introduces different swarm and evolutionary approaches to reduce number of multiplication needed to compute c. The exponent difficulty can be mapped directly in to an addition chain for computing exponent. As an alternative of multiplication, for a particular exponent value addition chain of sequence of integer can be generated using following properties.

- Value of the first element of the sequence always 1.
- Each consecutive element is generated by addition of two earlier elements.
- Last element value is identical as the exponent E. For example if we require to discover $p^{97}$ then one possible incompetent method to calculate (px , … ,xp) 97 times. Following competent method for exponent calculation is through addition series 1-2-3-5-10-20-40-50-90-95-97 that leads to the following scheme:

$a^1 = a$    $a^2 = a^1 a^1$    $a^3 = a^1 a^2$    $a^5 = a^2 a^3$    $a^{10} = a^5 a^5$    $a^{20} = a^{10} a^{10}$    $a^{40} = a^{20} a^{20}$    $a^{50} = a^{10} a^{40}$    $a^{90} = a^{40} a^{50}$    $a^{95} = a^5 a^{90}$    $a^{97} = a^2 a^{95}$

Where sequence length is 11. So, only 11 multiplications are needed to calculate $p^{97.}$ But for the similar example one more addition sequence can be generated with length 10 i.e. 1-2-4-6-10-20-24-48-96-97 which leads to the following sequence of computation.

$a^1 = a$    $a^2 = a^1 a^1$    $a^4 = a^2 a^2$    $a^6 = a^4 a^2$    $a^{10} = a^6 a^4$    $a^{20} = a^{10} a^{10}$    $a^{24} = a^{20} a^4$    $a^{48} = a^{24} a^{24}$    $a^{96} = a^{48} a^{48}$    $a^{97} = a^{96} a^1$.

So, from the example it is seen that for a particular exponent value there may be several addition sequence of different length.

For sinking number of multiplication sequence having tiny length for a particular exponent always be selected which is an optimization problem. There are more than a few deterministic [8] and stochastic [5] and heuristics based techniques proposed for solving this optimization problem.

This paper explored GA based stochastic methods to improve its capabilities in this search space.

The organization of this paper is as follows. Section 2 of this paper deals with problem statement. Some of the existing techniques proposed to handle this problem has been discussed in section 3. Proposed GADSA based strategy discussed in section 4. Complexity of the GADSA technique is discussed in section 5. Experimental results are given in section 6. Analysis regarding various aspects of the technique and results has been presented in section 7. Conclusions with the future scopes are drawn in section 8 and that of references at end.

## 2. PROBLEM DOMAIN

The problem deal by this paper is the optimization of addition sequence length X for a given exponent E. The aim is to find out addition sequence with minimum length (l). An addition sequence X with length l is defined as a sequence of positive integers X= $x_1$, $x_2$, $x_1$,…. $x_i$,…, $x_l$, with $x_1$=1, $x_2$ = 2 and $x_l$ = E, and $x_{i-1} < x_i < x_{i+1}$, where each $x_i$ is obtained by adding two previous elements $x_i = x_j + x_k$ with j, k< i for i>2.

## 3. RELATED WORK

A number of deterministic and probabilistic algorithms were proposed to obtain minimum addition chain. Some of them are Binary [3], m-ary [4], adaptive m-ary, Power tree, The Factor Method, Window, Adaptive Window, Artificial Immune system , GA [3, 4, 5, 9, 11, 12, 13].

Binary Method expands exponent to its binary version with length m followed by implementing a prearranged algorithm which scanned from left to right or right to left depending on the binary value of the scanned bit this algorithm computes fields squaring and multiplications function.
The technique has been enhanced by scanning more than one bit at a time. This new strategy is known as Window method [4] where k bits are scanned at a time. It is based on k-ary expansion of the exponent, where the bit of exponent are divided into k- bit words.

The Adaptive Window strategy (different versions are [3, 4, 5, 8, 9, 10]) can fine-tune its technique according to the definite form of the given exponent. Here binary inputs as exponent get divided into a series of variable having length zero and non zero digits, called window, which are processed. This algorithm is valuable for exponents with bit length more than 128 bits.

Despite the fact that most of the methods ate deterministic but a few probabilistic heuristics approaches were also proposed [11, 12, 13].

A simple GA based approach [11] with encoded addition chain as a chromosome was presented by N. Cruz-Cort´es et al. This GA based method uses simple selection strategy and one point crossover technique and applied on a small set of exponents and obtained a competitive result.

Artificial Immune System (AIS) [13] uses only feasible addition chains which works by emulating the Colonial Selection Principle where the unsurpassed individuals are cloned and these clones are also mutated.

## 4. PROPOSED GADSA TECHNIQUE

The objective of this technique is to obtain better performance compared to N. Cruz-Cortés and L. G. Osorio-Hernández [11, 12] which has a several drawbacks. Some of the drawbacks are:

- No intelligent mechanism has been used in generation of initial population. A rule for generation of gene has a worst time complexity because of linear search method in selection of valid random element in case of L. G. Osorio-Hernández method.

- Binary Tournament selection mechanism is not fitted for this problem because selection of best fitted chromosomes does not guarantee [12].

- Only one version of crossover has been proposed. For plenty miscellany of chromosomes use of only one crossover version may not produce fitted offspring with ample diversity in [11].

- During mutation only single mutant regenerated instead of multiple. So, no alternative to select finest mutant among several [11].

The shortcomings of existing GA based approach have been eliminated through GADSA approach. The criteria based on which existing GA based approach is modified are

- Introduction of Roulette Wheel selection mechanism as a replacement for Binary Tournament selection.

- Rule of generation of gene get customized by using binary search as an alternative of linear search to improve the time complexity.
- Variety of crossover techniques along with their arbitrary selection mechanism is also proposed.

- In the mutation phase N numbers of mutant get produced instead of single, from which best one get selected.

A schematic diagram of proposed GADSA has been represented using figure 1. At the first step GADSA technique creates an initial population using *Gene_Generation* ( ) method. Then using fitness function fitness of each chromosome is calculated and checks whether termination criteria are met or not. If termination criteria are not met then perform Roulette wheel selection procedure to select best chromosome to create mating pool. Different version of crossover depend on probability value is performed on the best fitted chromosomes followed by single point mutation technique. At last fitness of new chromosomes has been calculated using same fitness function used earlier and checks for termination criteria.

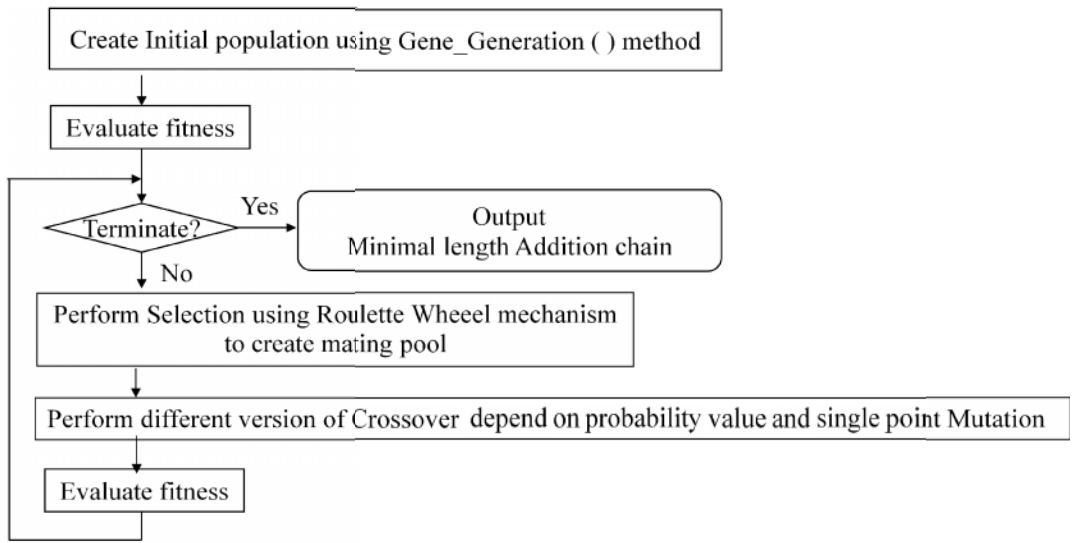

Figure 1.  Schematic diagram of proposed GADSA.

## 4.1 Encoding

In this GADSA invalid chromosomes get rule out. Only set of valid feasible chromosomes are consider for addition chain. Single dimension array of integer numbers represents a valid feasible addition chain which is consider as a chromosome of GA. Now, for exponent=43 following chromosomes shown in figure 2 can be structured which may have variable lengths.

Chromosome 1

| 1 | 2 | 3 | 6 | 12 | 24 | 36 | 42 | 43 |
|---|---|---|---|----|----|----|----|----|

Chromosome 2

| 1 | 2 | 4 | 8 | 16 | 32 | 36 | 38 | 39 | 43 |
|---|---|---|---|----|----|----|----|----|----|

Figure 2.  Example of valid chromosomes of variable length

## 4.2 Initial population

Initial population consists of group of valid chromosomes. Rules of generating each chromosome are as follows:

- *Double stepping: $x_i = 2x_{i-1}$ with 65% probability*
- *Addition: $x_i = x_{i-1} + x_{i-2}$ with 25% probability*
- *Addition of random gene with last gene: $x_i = x_{i-1} + rnd(x_0, x_{i-1} )$ with 10% probability where $x_{i-1} < x_i <=exponent$.*

Figure 3 shows invalid chromosome.

|   |   |   |   |    | *Random gene* |    | *Last gene* | *Invalid gene* |
|---|---|---|---|----|------|----|------|------|
| 1 | 2 | 3 | 6 | 12 | 24 | 36 | 42 | **66** |

Figure 3.  Example of invalid selection of random gene 24 which produces a new invalid gene 66 by adding last gene 42 where 66>43(exponent)

If random gene value is invalid (i.e. $x_i >$ exponent) then proposed technique checks diversity between exponent and last gene followed by binary search with the following criteria:

- If the desire gene value (difference result) is equal to the chain gene value then add the gene value with the last gene to generate the next gene (i.e. exponent).
- If the desire gene value is greater than the searched gene value then performs left traversal until a gene value of lower magnitude is found and add this lower magnitude value with last gene to generate the next gene.

First two genes value of any valid individual addition chain X are 1 followed by 2 and third gene value will be either 3 or 4 chosen randomly. Rest of the genes value gets generated through above stated rules. In this technique two unacceptable cases such as $x_i >$exponent and $x_i$ $x_{i-1}$ is not allowed. Algorithm for generating gene for construction of single chromosome is given below.

---

***Algorithm 1:  Gene_Generation (X, i, exponent)***

---

***Input:*** *A partial chromosome X = $x_1$, $x_2$, ..., $x_i$= e, where i represents the next position to be filled.*

***Output:*** *A feasible chromosome for a given exponent, with length l.*

***Method:*** *Each gene is generated by either double stepping or addition of last two genes or adding random gene with the last gene .Among these 3 methods one method is used each time depends on probability value.*

*Set m = i − 1*
*{/\*$x_m$ defines the last element of the chromosome \*/}*
 *while $xm$    exponent **do***
        ***if** prob(double)* ***then***
            *{/\*Applying Double Stepping )\*/}*
            *$x_m = 2x_m-1$*
        ***else if** prob(add)* ***then***
            *{/\*Selecting last genes for addition \*/}*
            *$x_m = x_m-1 + x_m-2$*
        ***else***
            *{/\*Selecting a random gene\*/}*
            *$x_m = x_m-1 + rnd(x_0, x_m-1)$*
        ***end if***
        ***while** $x_m > e$* ***do***
            *Look for a new feasible gene value, starting a binary search from last random gene value selected to first element if necessary.*
            ***end while***
            *m = m + 1*
***end while***

---

Algorithm for generating a complete chromosome with the help of Gene_Generation algorithm is given below.

---

***Algorithm 2:  Initial_Population (N )***

---

***Input:*** *Exponent and population size N.*
***Output:*** *A valid set of chromosome.*
***Method:***
***for** i=1 to N* ***do***
        *Set $x_1 = 1$ and $x_2 = 2$*
        *Allocate $x_3 = x_2 + rnd(x_1, x_2)$ where rnd() returns a random integer in the interval*
        *Generate a complete chromosome,*
        *(X, l) = **Gene_Generation** (X, 10, exponent)*
***end for***

---

## 4.3 Fitness Calculation

Fitness/objective function is associated with each chromosome. This indicates the degree of goodness of the encoded solution. The main aim of the fitness function is to select chromosomes having shorter length and by discarding the chromosomes having larger length. For example (1, 2, 4, 6, 8) has fitness value 4 i.e. one less than the actual length of the array.

## 4.4 Roulette Wheel Selection

The objective of Roulette Wheel selection is to select the good chromosomes among several chromosomes. In this mechanism fitness of chromosomes are plotted in a wheel and then spin the wheel. When the wheel stop then the probability of pointing arrow will be point in the zone having higher fitness is high also wheel zone having lesser fitness value will be selected with lower probability.

The aims of the selection mechanism are

- More copies to good strings
- Fewer copies to bad string
- Proportional selection scheme

    – Number of copies taken to be directly proportional to its fitness
    – Mimics the natural selection procedure to some extent.

Roulette wheel selection and Binary Tournament selection are two frequently used selection procedures.

In this proposed technique, Roulette wheel selection procedure is use to select best chromosome. In figure 4 best fitted chromosomes has greater partition space than worst so as to increase selection probability of better chromosome.

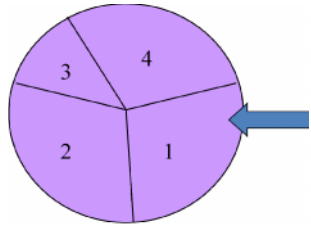

Figure 4. A Roulette wheel with several partitions according to fitness of chromosomes.

## 4.5 Crossover

In this proposed technique single point along with a two-point and uniform crossover techniques are used but the selection will depends on crossover probability of various techniques. In experimental result reveals that each of three crossover technique is suitable for different situations. Therefore, the technique focuses on choosing different crossover technique using crossover probability to produce good child chromosomes from the existing parents. In this technique, some of the abbreviations are used such as:

D= application of Double Rule,
A= application of Add Rule,
R= application of Random Rule,
G= application of *Gene_Generation* ( ) (Where corresponding parents rule is invalid).
E= Exchange of rules.

In single point crossover (with crossover probability 20%) the values into the parents' chromosome segments before the crossover point are copied into the corresponding offspring segments. For the chromosome segments after the crossover point the operator copies the rules (instead of the values) from the parents' segments into the offspring chromosomes.

In parent chromosome 1 in the figure 5 after the crossover point 6th, 7th, 8th and 9th genes are generated using Add, (A), Double (D), Random (R) and Random (R) rules respectively.

Parent Chromosome 1

| 1 | 2 | 3 | 6 (D) | **12 (D)** | 18 (A) | 36 (D) | 42 (R) | 43 (R) |
|---|---|---|---|---|---|---|---|---|
| 1 | 2 | 3 | 4 | 5 | 6 | 7 | 8 | 9 |

Figure 5. Example of parent chromosome 1 in single point crossover.

In parent chromosome 2 in the figure 6 after the crossover point 6th, 7th, 8th , 9th and 10th genes are generated using Double (D), Random (R) , Random (R), Random (R) and random (R) rules respectively.

Parent Chromosome 2

| 1 | 2 | 4 | 8 (D) | **16 (D)** | 32 (D) | 36 (R) | 38 (R) | 39 (R) | 43 (R) |
|---|---|---|---|---|---|---|---|---|---|
| 1 | 2 | 3 | 4 | 5 | 6 | 7 | 8 | 9 | 10 |

Figure 6. Example of parent chromosome 2 in single point crossover.

For this example consider crossover point = 5. After single point crossover gene value of parent chromosomes 1 and 2 will be copied to the corresponding child chromosomes 1 and 2 respectively up to crossover point. For generating rest of the genes after the crossover point in child chromosome1 rule used in parent chromosome 2 is follow i.e. Double (D), Random (R) , Random (R) and Random (R) for gene position at 6th, 7th, 8th and 9th respectively (as shown in figure 7).

Child Chromosome 1

| 1 | 2 | 3 | 6 | **12** | 24 (D) | 30 (R) | 42 (R) | 43 (R) |
|---|---|---|---|---|---|---|---|---|
| 1 | 2 | 3 | 4 | 5 | 6 | 7 | 8 | 9 |

Figure 7. Child chromosome 1 after single point crossover with crossover point p=5.

In child chromosome 2 , for generating rest of the genes after the crossover point, rule used in parent chromosome1 is follow i.e. Add (A) for gene at position 6th then for position 7th instead of Double rule, *Gene_Generation ( )* (G) rule is use. In position 8th and 9th Random rule (R) is used (as shown in figure 8).

Child Chromosome 2

| 1 | 2 | 4 | 8 | **16** | 24 (A) | 40 (G) | 42 (R) | 43 (R) |
|---|---|---|---|---|---|---|---|---|
| 1 | 2 | 3 | 4 | 5 | 6 | 7 | 8 | 9 |

Figure 8. Child chromosome 2 after single point crossover with crossover point p=5.

In two point crossover (with crossover probability 35% ) the values into the parents' chromosome segments before the first crossover point are copied into the corresponding offspring segments. For the chromosome segments after the first ($p$) and second ($q$) crossover points, the operator copies the rules (instead of the values) from the parents' segments into the offspring chromosomes.

In parent chromosome 1 after the first crossover point at $4^{th}$ position and before the second crossover point at $8^{th}$ position genes are generated using Add (A), Random (R) and Add (A) rules at $5^{th}$, $6^{th}$ and $7^{th}$ position respectively. After second crossover point at $8^{th}$ position genes are generated using Random (R) and Random (R) rules for the position $9^{th}$ and $10^{th}$ respectively (as shown in figure 9).

Parent Chromosome 1

| 1 | 2 | 3 | **6** **(D)** | 9 (A) | 12 (R) | 18 (A) | **36** **(D)** | 42 (R) | 43 (R) |
|---|---|---|---|---|---|---|---|---|---|
| 1 | 2 | 3 | 4 | 5 | 6 | 7 | 8 | 9 | 10 |

Figure 9.  Example of parent chromosome 1 in two point crossover.

In parent chromosome 2 after the first crossover point at $4^{th}$ position and before the second crossover point at $8^{th}$ position genes are generated using Double (D), Add (A) and Add (A) rules at $5^{th}$, $6^{th}$ and $7^{th}$ position respectively. After second crossover point at $8^{th}$ position genes are generated using Random (R), Random (R) and Random (R) rules for the position $9^{th}$, $10^{th}$ and $11^{th}$ respectively (as shown in figure 10).

Parent Chromosome 2

| 1 | 2 | 4 | **5** **(R)** | 10 (D) | 15 (A) | 25 (A) | **35** **(A)** | 40 (R) | 42 (R) | 43 (R) |
|---|---|---|---|---|---|---|---|---|---|---|
| 1 | 2 | 3 | 4 | 5 | 6 | 7 | 8 | 9 | 10 | 11 |

Figure 10.  Example of parent chromosome 2 in two point crossover.

In child chromosome 1 after the first crossover point at $4^{th}$ position and before the second crossover point at $8^{th}$ position genes are generated using the same rules that were used in parent chromosome 2 i.e. Double (D) ,Add (A) and Add (A) rules at $5^{th}$, $6^{th}$ and $7^{th}$ position respectively. After second crossover point at $8^{th}$ position genes are generated using same rule as parent chromosome 2 i.e. Random (R) rules for the $9^{th}$ position (as shown in figure 11).

Child Chromosome 1

| 1 | 2 | 3 | **4** **(R)** | 8 (D) | 12 (A) | 20 (A) | **40** **(D)** | 43 (R) |
|---|---|---|---|---|---|---|---|---|
| 1 | 2 | 3 | 4 | 5 | 6 | 7 | 8 | 9 |

Figure 11.  Child chromosome 1 after two point crossover with crossover point p=4 and q=8.

In child chromosome 2 after the first crossover point at $4^{th}$ position and before the second crossover point at $8^{th}$ position genes are generated using the same rules that were used in parent chromosome 1 i.e. Add (A), Random (R) and Add (A) rules at $5^{th}$, $6^{th}$ and $7^{th}$ position respectively. After second crossover point at $8^{th}$ position genes are generated using the same rules that were used in parent chromosome 1 i.e. Random (R) and Random (R) rules for the position 9[th] and10[th] respectively (as shown in figure 12).

Child Chromosome 2

| 1 | 2 | 4 | **8** **(D)** | 12 (A) | 16 (R) | 28 (A) | **40** **(G)** | 42 (R) | 43 (R) |
|---|---|---|---|---|---|---|---|---|---|
| 1 | 2 | 3 | 4 | 5 | 6 | 7 | 8 | 9 | 10 |

Figure 12. Child chromosome 2 after two point crossover with crossover point p=4 and q=8.

Uniformed Crossover (with crossover probability 45%) In this type a mask is used for performing crossover operation, where the mask size is equal to the chromosome size and mask contains randomly generated binary values. Two rules for generating offspring's are:

- 1[st] and 2[nd] position value of both chromosomes get unchanged and 3[rd] position value gets exchanged with other chromosome if corresponding mask value is 1. Otherwise no change.

- For a particular position if mask value is 1 then rules of the parent chromosomes for this position gets exchanged to generate the corresponding value in the child. Otherwise there is no exchanged of rules child has to follow corresponding parent's rule to generate value of a particular position.

In parent chromosome 1 in the figure 13 genes at 4[th], 5[th],6[th], 7[th], 8[th] ,9[th] and 10[th] position are generated using Double (D), Add, (A),Random (R), Add (A), Double (D), Random (R) and Random (R) rules respectively.

Parent Chromosome 1

| 1 | 2 | 3 | 6 (D) | 9 (A) | 12 (R) | 18 (A) | 36 (D) | 42 (R) | 43 (R) |
|---|---|---|---|---|---|---|---|---|---|
| 1 | 2 | 3 | 4 | 5 | 6 | 7 | 8 | 9 | 10 |

Figure 13. Example of parent chromosome 1 in uniform crossover.

In parent chromosome 2 genes at 4[th], 5[th],6[th], 7[th], 8[th] ,9[th] , 10[th] and 11[th] position are generated using Random (R), Double (D), Add, (A), Add, (A), Add, (A), Random (R), Random (R) and Random (R) rules respectively (as shown in figure 14).

Parent Chromosome 2

| 1 | 2 | **4** | **5** **(R)** | 10 (D) | 15 (A) | 25 (A) | **35** **(A)** | 40 (R) | 42 (R) | 43 (R) |
|---|---|---|---|---|---|---|---|---|---|---|

Figure 14. Example of parent chromosome 2 in uniform crossover.

Mask is used for indicating the gene positions where the uniform crossover will take place.

Mask

| 1 | 0 | 1 | 0 | 1 | 1 | 0 | 0 | 1 | 1 |
|---|---|---|---|---|---|---|---|---|---|
| 1 | 2 | 3 | 4 | 5 | 6 | 7 | 8 | 9 | 10 |

Figure 15.  Mask for uniform crossover, where crossover will take place at the positions having value 1

Mask values are set at $1^{st}$, $3^{rd}$, $5^{th}$, $6^{th}$, $9^{th}$ and $10^{th}$ positions. In child chromosome 1 and 2 the value of the gene at $1^{st}$ and $3^{rd}$ position get exchanged with parent chromosome 1 and 2.  For the position $5^{th}$, $6^{th}$ and $9^{th}$ and $10^{th}$ the rules for gene generation of parent chromosome 1 and chromosome 2 get exchanged for generation gene values in corresponding child (as shown in figure 16 and 17).

Child Chromosome 1

| 1 (E) | 2 | 4 (E) | 8 (D) | 16 (E) | 24 (E) | 36 (A) | 40 (G) | 42 (E) | 43 (E) |
|---|---|---|---|---|---|---|---|---|---|
| 1 | 2 | 3 | 4 | 5 | 6 | 7 | 8 | 9 | 10 |

Figure 16.   Child chromosome 1 after uniform crossover with crossover at the position $1^{st}$, $3^{rd}$, $5^{th}$, $6^{th}$, $9^{th}$ and $10^{th}$.

Child Chromosome 2

| 1 (E) | 2 | 3 (E) | 8 (D) | 12 (E) | 16 (E) | 28 (A) | 40 (G) | 42 (R) | 43 (R) |
|---|---|---|---|---|---|---|---|---|---|
| 1 | 2 | 3 | 4 | 5 | 6 | 7 | 8 | 9 | 10 |

Figure 17. Child chromosome 2 after uniform crossover with crossover at the position $1^{st}$, $3^{rd}$, $5^{th}$, $6^{th}$, $9^{th}$ and $10^{th}$.

## 4.6 Mutation

In this proposed scheme N numbers of mutants are generated instead of just one [18]. One mutation point is arbitrarily selected in each offspring and the segment before the mutation point is copied *N* times. The function ***Gene-Generation*** **( )** is use to generate gene for the *N* individuals. Only the best new mutated child will be allowed to survive, replacing the original child.

A example is given in figure 18 for the exponent = 97, where a mutation point (the position with value 27) is selected into the original child then all the gene values up to the mutation point are copied   into *N* = 4 mutated chromosomes. The remaining genes value ogf each mutated chromosomes are computed by using the function *Gene_Generation* ( ). In this example the best resulting mutated chromosome is the second one meanwhile the worst is the one and fourth one. Only the best child will be allowed to survive by replacing the original one (as illustrated in figure 18 and figure 19).

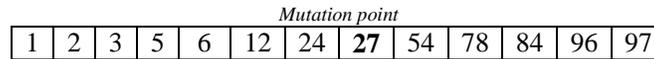

Figure 18.  Chromosome with mutation point at 8<sup>th</sup> position.

In figure 19 different mutants are produced from one single chromosome.

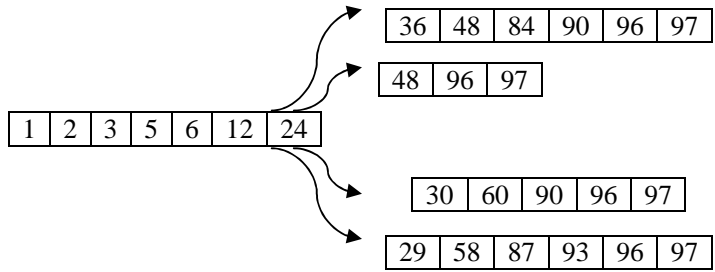

Figure 19. Generations of different mutants

---

**Algorithm 3:**   *Mutation (Child)*

---

**Input:** *An Individual Child = (X, l).*
**Output:** *The Best Mutated Child Version Xm = (Xm, l)*
**Method:** *After choosing a mutation point all the genes up to the  mutation point are copied to N mutated chromosomes then Gene Generation ( ) is used to generate rest of the genes of each mutant after mutation point.*

**If** *Prob(Pm)* **then**
   *Generate N mutants of a chromosome.*
*{/*Select a common limit to generate next mutation point*/}*
   *i = rnd(3, l(Child))*
   **for** *a = 1 to N* **do**
        *Select a random position Child$_{a,b}$, where*
        *b = rnd(0, i−1) is a random element of the actual clone.*
        *Generate a new element for Child$_{a,i}$+1 (mutation point),such that*
        *Child$_{a,i}$ +1 = Child$_{a,i}$ + Child$_{a,b}$*
        *Repair addition chain starting from new generated element*
        *Child$_a$ = Gene_Generation(Child$_{a,i}$ + 1, exponent)*
   **end for**
   *Choose the best mutation version.*
**end if**

---

**Algorithm 4:   GADSA based Optimal Addition Chain (exponent)**

---

**Require:** *POPULATION_SIZE is odd.*
**Input:** *An exponent*
**Output:** *An optimal addition chain X = x$_1$, x$_2$, ..., x$_l$= exponent*
**Method:** *Initial population is constructed using InitialPopulation( )then using Roulette Wheel selection 2 parent chromosomes are selected after that crossover and mutation process is used to generate optimal addition chain.*

.
*{/*Setting up initial population*/}*

*for* $i$ = 1 to POPULATION_SIZE *do*
    *Parents$_i$* = *InitialPopulation(exponent)*
*end for*
*for* $i$ = 1 to MAXIMUM_GENERATION *do*
    *Randomize parents population*
    *k = 0*
    *while* $k$ < (POPULATION_SIZE − 1) *do*
        *{/\*Select a couple of parents (P1, P2)\*/}*
        *P1 = selection(T)*
        *P2 = selection(T)*
        *{/\*Generate two new children ($C_k$, $C_k$+1), by applying*
        *crossover operator\*/}*
        *crossover(P1, P2)*
        *{/\*Apply the mutation operator to each new created child \*/}*
        *Mutation(Child$_k$)*
        *Mutation(Child$_k$ +1)*
        *Set k = k +2*
    *end while*
  *Replace parent's population with children population*
  *end for*
*Report fittest individual*

## 5. COMPLEXITY ANALYSIS OF GADSA TECHNIQUE

**Step 1.** *Gene_Generation* ( ) method is used to generate genes for chromosome construction. In this method either of 3 techniques are use to generate genes. In addition of last two genes and double stepping methods take constant amount of time to perform arithmetic operation. Selecting random gene for addition methods use binary search technique to search required gene value from the chain using O (log n) amount of time.

**Step 2.** GA are not chaotic, they are stochastic. The complexity depends on the genetic operators, their implementation (which may have a very significant effect on overall complexity), the representation of the individuals and the population, and obviously on the fitness function. Given the usual choices (Single point mutation, one point crossover, two point crossover, uniform crossover and roulette wheel selection) a GA complexity is O(g(nm + nm + n)) with g the number of generations, n the population size and m the size of the individuals. Therefore the complexity is on the order of O(gnm)).

## 6. EXPERIMENTAL RESULTS

Performance of GADSA based approach is compared with its previous version [11, 12, 14, 15] and other deterministic and heuristic based [3, 4, 5, 9, 13] approaches with the aspire to (1) show that performance is competitive (or even improved) with those provided by other heuristic-based or deterministic approaches and (2) to solve even more complex instances of the problem. Complexity of the problem depends on the value of exponent. This section highlighted on the best solution reached so far i.e. quality and statistical analysis for reviewing consistency. Different parameters values that are used in proposed GADSA based approaches are obtained by a trial-and-error method by favouring the best overall performance.

## 6.1 Parameters used in GADSA

Population Size (*POPULATION_SIZE*) = 200, maximum number of generations (*MAXGEN*) = 300, single-point crossover with the crossover probability=20%, two-point crossover with the crossover probability=35%, uniform crossover with the crossover probability=45% , crossover Rate = 0.4%, mutation Rate = 1.0%, rate of double stepping=0.65, rate of previous position addition =0.25, rate of addition of random element with last element= 0.1

The first set of experiments compute the total accumulated addition chains for a fixed set of exponents. An accumulated addition chain for a maximum value *P*, represents the sum of all addition chains obtained for all the exponents *1, 2, . . .,P* as stated in following equation. An accumulated addition chain for a given exponent represents the sum of all addition chains obtained for each number in the sequence defined by exponent stated in equation (2).

$$Accumulated\_Add\_Chain = \sum_{i=1}^{P} Optimal\_Add\_Chain \qquad (2)$$

A smaller value of Accumulated_Add_Chain represents better performance by the algorithm.

In table 1 set of accumulated addition chains for all exponents less than: 512 ($e \in$ [1, 512]), 1000 ($e \in$ [1, 1000]), 2000 ($e \in$ [1, 2000]), 2048 ($e \in$ [1, 2048]) and 4096 ($e \in$ [1, 4096]), is presented in order to compare the results with respect to earlier [11, 12, 14, 15] and proposed GADSA and some other heuristic [13] and deterministic [3, 4, 5, 9] algorithms that has been proposed so far.

Table 1 represents the optimal value and the best results reached by proposed GADSA and existing heuristics and deterministic approaches. From this table it is observed that proposed GADSA performs better than GA [11], Quaternary, Binary technique at [[1,512]. In the range of [1, 1000] GADSA reaches the optimal value and performs better than GA [11], REPLS-GA [12], PSO [14], AIS [13]. In the range of [1, 1024], [1, 2000], [1, 2048], [1, 4096] proposed GADSA has obtained the best results compare to the other existing methods.

Table 1
Comparison of best results obtained by the former version of (GA [11, 12], PSO [14]), AIS [13],
Quaternary, Binary and proposed GADSA

| Exponent | Optimal | GADSA | GA [11] | REPLS-GA [12] | PSO [14] | AIS [13] | Quaternary | Binary |
|---|---|---|---|---|---|---|---|---|
| [1,512] | 4924 | **4924** | 4925 | 4924 | 4924 | 4924 | 5226 | 5388 |
| [1,1000] | 10808 | **10808** | 10818 | 10809 | 10813 | 10813 | 5603 | 5812 |
| [1,1024] | 11115 | **11117** | - | - | 11120 | 11120 | 11862 | 12301 |
| [1,2000] | 24063 | **24072** | 24124 | 24076 | 24095 | 24108 | 25923 | 26834 |
| [1,2048] | 24731 | **24739** | - | 24748 | 24765 | 24778 | 26664 | 27662 |
| [1,4096] | 54425 | **54472** | - | 54487 | 55609 | 56617 | 58678 | 61455 |

In table 2 statistical results obtained in 40 independent runs for all exponent set consider so far by the proposed GADSA are listed. Binary method, adaptive window method or other non-deterministic heuristics methods [3, 4, 5, 9, 11, 12, 13] are not be able to obtained shortest addition chain for optimizing a given exponent. So a given exponent is hard to optimize using these strategies. Whereas proposed GADSA have the capabilities to address and solve this problem.

Table 2
Statistical Results of Proposed GADSA

| Exponent | Best | Average | Median | Worst |
|----------|------|---------|--------|-------|
| [1,512] | 4924 | 4924.02 | 4924 | 4925 |
| [1,1000] | 10808 | 10809.76 | 10810 | 10813 |
| [1,1024] | 11117 | 11118.34 | 11118.5 | 11120 |
| [1,2000] | 24072 | 24078.49 | 24079 | 24085 |
| [1,2048] | 24739 | 24745.73 | 24746 | 24751 |
| [1,4096] | 54472 | 54484.91 | 54485 | 54497 |

Table 3 shows average length of addition chain for proposed GADSA vs. Binary, Quaternary technique where GADSA obtained shortest addition chain in exponent size compare to the Binary and Quaternary techniques.

Table 3
Average length of addition chain for proposed GADSA vs. Binary, Quaternary method

| Exponent size | GADSA | Binary | Quaternary |
|---------------|-------|--------|------------|
| 32 | 40 | 47 | 43 |
| 64 | 77 | 95 | 87 |
| 128 | 155 | 191 | 175 |
| 256 | 311 | 383 | 351 |
| 512 | 617 | 767 | 703 |
| 1024 | 1270 | 1535 | 1407 |

Table 4 shows addition chains for some of the special class of exponent. Binary, Quaternary and some other existing techniques are not able to generate minimal length addition chain for this class of exponents where as proposed GADSA can generate minimal length addition chain for this class of exponents with addition chain length 27.

Table 4
Shortest addition sequence for a special class of exponents

| Exponent | Addition Chain | GADSA |
|----------|----------------|-------|
| 3704431 | 1 2 4 5 9 18 36 72 144 288 576 1157 2314 4628 9256 18512 37024 74048 148096 296192 592384 1184768 1185349 2370698 3556047 3704143 3704431 | 27 |
| 3922763 | 1 2 3 6 12 24 26 52 104 208 416 832 1664 3328 3331 6659 9990 19980 39960 79920 159840 163171 326342 652684 1305368 1958052 3916104 3922763 | 27 |
| 2948207 | 1 2 3 4 7 14 28 29 58 116 232 239 478 956 1912 3824 3853 7677 15354 30708 61416 122832 245664 491328 982656 1965312 2947968 2948207 | 27 |
| **3093839** | 1    2    3    5    10    20    30    60    120    150    151    302    604    1208    2416    4832    9664    19328    38656    77312    154624    309248    618496    1236992    2473984    3092480    3093688    3093839 | 27 |
| **3243931** | 1    2    4    8    16    32    64 27    128    256    258    514    515    1029    2058    4116    8232    16464    32928    65856    66371    132227    198083    396166    792332    1584664    3169328    3235699    3243931 | 27 |
| **3325439** | 1    2    4    8    16    17    33    66    132    264    528    1056    2112    4224    4241    8482    16964    33928    67856    135712    271424    271457    542914    1085828    1085861    2171722    3257583    3325439 | 27 |

## 7. ANALYSIS OF RESULTS

In larger range of exponent value i.e. [1-2000], [1-2048], [1-4096] GADSA outperforms than existing GA [11, 12]. GADSA performs better than REPLS_GA [12] and simple GA [11] in the exponent range of [1-1000], [1-2000], [1-2048], [1-4096]. This proposed GADSA performs better than AIS [13], Quaternary and Binary [3, 4, 5] techniques in the entire considered exponent range. Overall findings suggest that proposed GADSA performs outstandingly well among all the techniques and its results close to optimal values in most cases.

## 8. FUTURE SCOPE & CONCLUSION

In this paper several modifications of existing GA [11, 12] techniques has been done for finding the optimal addition sequence for a given exponent. Gene generation method, searching techniques for fittest value in chromosome creation, crossover techniques are get modified through proposed GADSA. This proposed method is also compared with the existing heuristics and deterministic approaches [3, 4, 5, 11, 12, 13, 14, 15]. GADSA outperformed all techniques mostly in larger exponent i.e. [1-2000], [1-2048], [1-4096]. Furthermore, proposed version of heuristic based approaches able to generate optimal addition sequence on exponent where no other heuristic based approach has reported results in the specialized literature.

Future scope of the proposed GADSA is to apply other soft computing tools in this problem domain and to design better mutation technique in proposed GADSA approach. Another future scope is to analyze the parameters required by these algorithms to reduce the number of evolutions without affecting the efficiency and test the algorithms with larger exponent (more than 160 bits).

## ACKNOWLEDGEMENTS

The author expresses deep sense of gratitude to the DST, Govt. of India, for financial assistance through INSPIRE Fellowship leading for a PhD work under which this work has been carried out.

## Authors

Arindam Sarkar

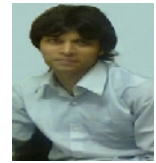

INSPIRE Fellow (DST, Govt. of India), MCA (VISVA BHARATI, Santiniketan, University First Class First Rank Holder), M.Tech (CSE, K.U, University First Class First Rank Holder). Total number of publications 8.

Jyotsna Kumar Mandal

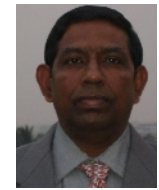

M. Tech.(Computer Science, University of Calcutta), Ph.D.(Engg., Jadavpur University) in the field of Data Compression and Error Correction Techniques, Professor in Computer Science and Engineering, University of Kalyani, India. Life Member of Computer Society of India since 1992 and life member of cryptology Research Society of India. Dean Faculty of Engineering, Technology & Management, working in the field of Network Security, Steganography, Remote Sensing & GIS App lication, Image Processing. 25 years of teaching and research experiences. Eight Scholars awarded Ph.D. one submitted and 8 are pursuing. Total number of publications 230.